\documentclass[iop]{emulateapj}
\usepackage{epsfig}

\slugcomment{Accepted for publication in the Astrophysical Journal}

\shorttitle{The Orbital Period of Sco X-1}
\shortauthors{Hynes \& Britt}

\begin{document}

\title{The Orbital Period of Scorpius X-1}

\author{Robert I. Hynes\altaffilmark{1}, Christopher T. Britt}
\affil{Louisiana
  State University, Department of Physics and Astronomy, 202 Nicholson
  Hall, Tower Drive, Baton Rouge, LA 70803, USA}

\altaffiltext{1}{E-mail: rih@phys.lsu.edu} 

\begin{abstract} 

The orbital period of Sco~X-1 was first identified by
\citet{Gottlieb:1975a}.  While this has been confirmed on multiple
occasions, this work, based on nearly a century of photographic data,
has remained the reference in defining the system ephemeris ever
since.  It was, however, called into question when
\citet{Vanderlinde:2003a} claimed to find the one-year alias of the
historical period in {\it RXTE}/ASM data and suggested that this was
the true period rather than that of \citet{Gottlieb:1975a}.  We
examine data from the All Sky Automated Survey (ASAS) spanning
2001--2009.  We confirm that the period of \citet{Gottlieb:1975a} is
in fact the correct one, at least in the optical, with the one-year
alias strongly rejected by these data.  We also provide a modern time
of minimum light based on the ASAS data.

\end{abstract}

\keywords{Stars: Individual: Sco X-1 --- X-rays: Binaries --- Binaries: Close}

\section{Introduction}

The first extra-Solar X-ray source discovered was the low-mass X-ray
binary Sco~X-1 \citep{Giacconi:1962a}.  Its optical counterpart,
V818~Sco, was discovered by \citet{Sandage:1966a}, paving the way for
many subsequent multiwavelength studies.  The binary period is widely
accepted to be 18.9\,hr based on the discovery of a photometric
modulation by \citet{Gottlieb:1975a} and spectroscopic confirmation by
\citet{Cowley:1975a}.  We now know that Sco~X-1 contains a low-mass
late-type donor transferring mass onto a neutron star at a rather high
rate.  The modulation arises from X-ray heating of the donor star,
which also manifests as narrow emission lines of N\,{\sc iii} and
C\,{\sc iii} moving in phase with the donor star
\citep{Steeghs:2002a}.

\citet{Gottlieb:1975a} obtained the period of
$0.787313\pm0.000001$\,days quite remarkably by examining archival
photographic plates from 1889 to 1974.  A sinusoidal modulation of
full amplitude around 0.2--0.3\,mag was found in several independent
datasets, with considerable scatter around the mean curve
\citep{Gottlieb:1975a,Wright:1975a}.  While the long baseline of
photographic observations defined the period to incredible precision,
the sparse sampling left a plethora of aliases, and
\citet{Gottlieb:1975a} identified strong signals at one-day,
one-month, and one-year aliases of their favored period.  Of these,
the one-year alias has been by far the hardest to reject.  Several
subsequent photometric studies reproduced the modulation, but none
improved the ephemeris, or resolved the one-year alias issue
\citep{vanGenderen:1977a,Augusteijn:1992a}

Spectroscopic confirmation of this period was suggested by
\citet{Gottlieb:1975a} and \citet{Wright:1975a}, and demonstrated
conclusively by \citet{Cowley:1975a}, who found  a
period of $0.787\pm0.006$\,days, and again by
\citet{LaSala:1985a}.  Both of these works performed a period search on
the data, but in both cases the frequency resolution was limited by
only observing over a baseline of a week.  Other spectroscopic
analyses of these and other data have also found variations at this
period, \citep{Crampton:1976a,Bord:1976a,Steeghs:2002a}, but no other
groups have performed a rigorous independent period search.

Several groups also searched for the orbital period in X-ray data,
with initially no success
\citep{Holt:1976a,Coe:1980a,Priedhorsky:1987a,Priedhorsky:1995a}.  The
only positive detection of an orbital period in X-rays came from
\citet{Vanderlinde:2003a} based on a multi-year {\it RXTE}/ASM
dataset.  They did not find exactly the \citet{Gottlieb:1975a} period,
but instead the one-year alias (0.78893\,days) with a modulation
around 1\,\%.  Given the intensive multi-year coverage of {\it RXTE}
this is surprising, since this dataset should not be susceptible to
the one-year alias problem.  \citet{Vanderlinde:2003a} therefore
claimed that their period was the true orbital period and that
\citet{Gottlieb:1975a} had misidentified the alias.  While this result
was tantalizing, \citet{Levine:2011a} could not reproduce this period
using a larger {\it RXTE} dataset.  They did, however, not use as
sophisticated an analysis as \citet{Vanderlinde:2003a}, leaving open
the possibility that the X-ray period could be real.

Surprisingly, then, fifty years after discovery of the prototypical
LMXB Sco~X-1, there remain doubts about its most fundamental
parameter, the orbital period.  While the original optical ephemeris
of \citet{Gottlieb:1975a} has remained the standard reference for the
37\,years since its publication, it remains to be resolved whether
this, or the X-ray period of \citet{Vanderlinde:2003a}, is the true
orbital period.  To attempt to resolve these questions, and update the
ephemeris of Sco~X-1 with modern data, we examine here archival
photometry from the All Sky Automated Survey (ASAS).  This nine year
dataset has both the long baseline to determine a precise period, and
coverage of a large enough fraction of a year to finally break the
one-year alias problem using optical data.

\section{Observations}
\label{DataSection}

The All Sky Automated Survey (ASAS) monitored Sco~X-1 from 2001 to
2009 \citep{Pojmanski:2002a}.  We note that while Sco~X-1 was not
included in the ASAS Catalog of Variable Stars (ACVS) its photometry
is in the ASAS-3 Photometric $V$ Band Catalog in two datasets,
161955--1538.4 and 161955--1538.5.  The Sco X-1 datasets include 640
observations from 2001 January 22 to 2009 October 5.  With multiyear
coverage spanning typically about 270\,days of the year, it is ideally
suited for obtaining an updated ephemeris and breaking the one-year
alias.

We performed our analysis for a range of choices of data grades and
apertures to optimize our filter criteria.  For final analysis, we
retained the 567 grade A or B observations, and used the smallest ASAS
aperture.  Inclusion of grade C or worse data, or use of larger
aperture data, significantly reduced the quality of the fits.

\section{Ephemeris}

\label{PeriodSection}

\begin{figure}
\includegraphics[angle=90,width=3.5in]{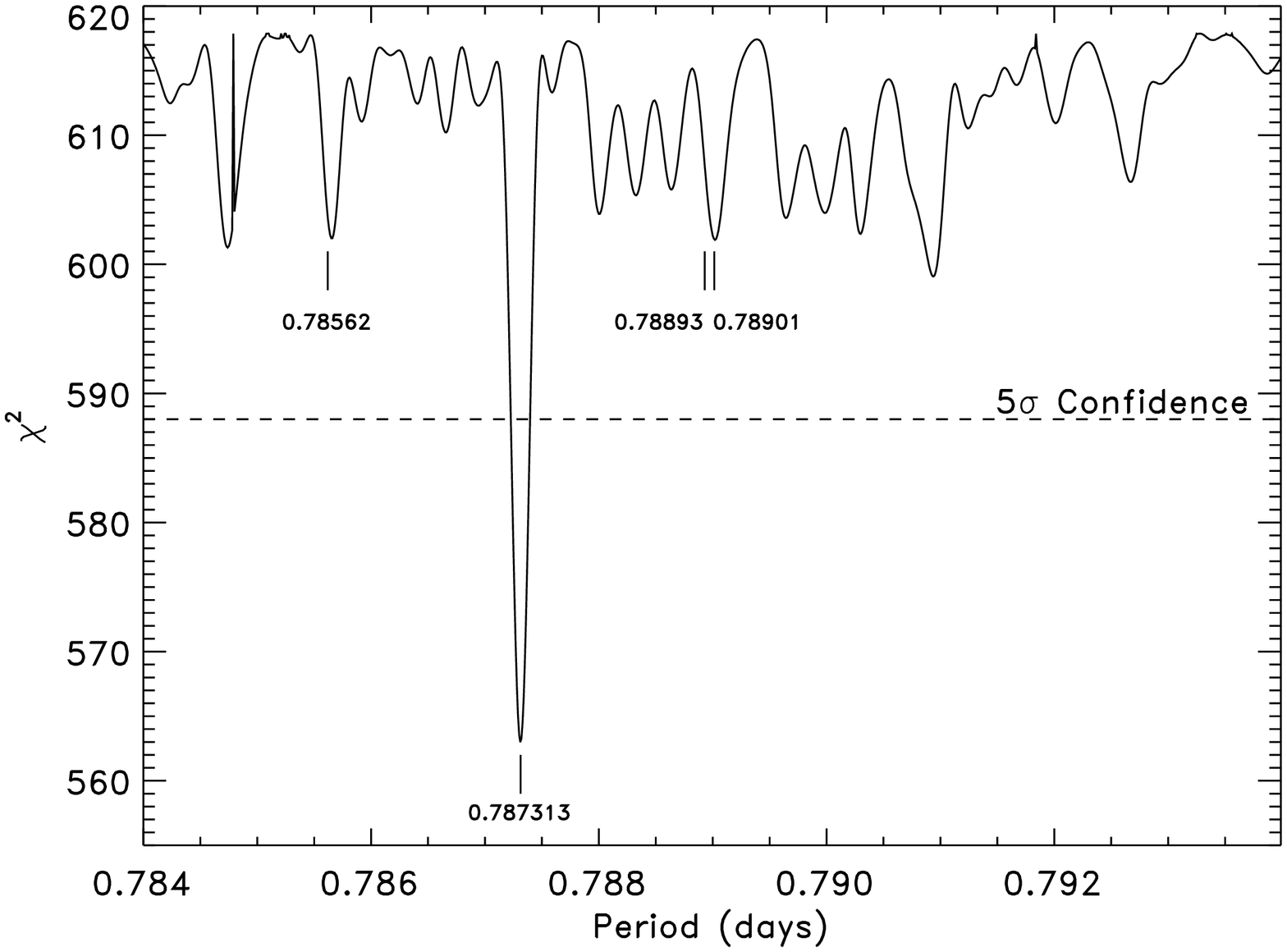}
\caption{$\chi^2$ as a function of trial period for sinusoidal fits to
  ASAS data.  We show the \citet{Gottlieb:1975a} period of
  0.787313\,days and the \citet{Vanderlinde:2003a} period of
  0.78893\,days for comparison.  We also show calculated one-year
  aliases of the preferred period at 0.78562 and 0.78901\,days.  The
  \citet{Gottlieb:1975a} period is strongly favored by the ASAS data.  While
  some signal is seen at the alternative periods, all are rejected at
  greater than 5-$\sigma$ confidence.}
\label{PeriodFig}
\end{figure}

To determine the orbital period we performed a sinuoidal fit to the
data points.  Since the scatter around the model is dominated by
intrinsic flickering rather than photometric uncertainties, we
assigned a mean uncertainty of 0.30\,mag to each point to represent
the flickering.  This was chosen to produce a minimum $\chi^2$ equal
to the number of degrees of freedom.  We then evaluated sinusoidal
fits over a range of trial periods.  For each period the best-fitting
mean magnitude, amplitude, and phasing were determined using the
downhill simplex algorithm \citep{Nelder:1965a}.  We show the results
in the vicinity of the disputed periods in Fig.~\ref{PeriodFig}.

We see that the \citet{Gottlieb:1975a} period is reproduced exactly to
within the limits of our frequency resolution.  Our formal best period
is $0.787313\pm0.000015$\,days.  The uncertainty quoted is a formal
1-$\sigma$ error determined from the $\Delta\chi^2=1$ confidence range
in period.  We verified the uncertainty using the bootstrap method
with 30 resamplings of the data.  This gave a consistent 1-$\sigma$
uncertainty ($1.6\times10^{-5}$).  We also show the period of
\citet{Vanderlinde:2003a}, and the one-year aliases with which they
associated it.  We find that none of these alternatives are consistent
with the ASAS data, and all can be rejected at better than 5-$\sigma$
confidence.  We therefore cannot directly improve on the period of
\citet{Gottlieb:1975a} using the ASAS data, which is not surprising as
that used data drawn from nearly a hundred year baseline.  We can,
however, overcome the limitation of that dataset in its vulnerability
to one-year aliases, as the ASAS data has much wider coverage within a
year.

Using the same $\chi^2$ approach, we determine a mean time of minimum
of $2453510.329\pm0.017$.  This corresponds to an offset of very close
to 17057 cycles from the time of minimum of \citet{Gottlieb:1975a}.
If we project their time of minimum forwards we predict
$2453510.328\pm0.024$, with equal contributions to the uncertainty
from their time of minimum (quoted as 0.022 cycles) and their period
($10^{-6}$\,days).  Our time of minimum is completely consistent with
theirs (a remarkable testament to the accuracy of their historical
ephemeris), but at this point our modern measurement of the time is somewhat
better constrained for use with modern data.

\begin{figure}
\includegraphics[angle=90,width=3.5in]{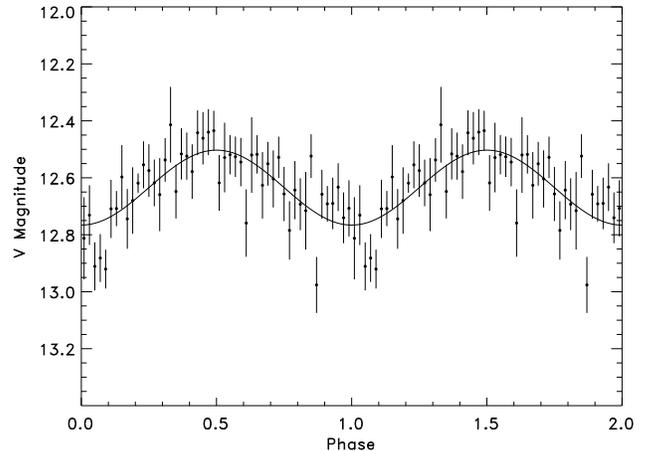}
\caption{Folded and phase-binned ASAS lightcurve of Sco~X-1.  The
  data have been grouped into 50 phase bins and plotted twice.
  Errorbars are empirical and indicate the error on the mean of each
  bin.  The model plotted is the best fitting sine wave determined
  Section~\ref{PeriodSection}.}
\label{LightcurveFig}
\end{figure}

Finally, we show in Figure~\ref{LightcurveFig} the ASAS lightcurve folded on
our derived ephemeris, together with the best fitting sine wave.  The
mean $V$ band brightness is 12.63, and the full-amplitude is 0.26\,mag,
comparable to that found by \citet{Gottlieb:1975a} and
\citet{Wright:1975a}.

\section{Discussion}

We have established that in optical photometry the 0.787313\,day
period produces a stable modulation over 120\,years of observation.
The ephemeris of \citet{Gottlieb:1975a} reliably and precisely
predicts the time of minimum in the ASAS data, over 17,000
intervening cycles.  It is hard to imagine any clock other than the
orbital period providing this stability.  This has to be the true
orbital period.

The question then arises as to what, if anything,
\citet{Vanderlinde:2003a} detected.  We of course should allow that it
was a spurious detection, until it can be reproduced with data from
the remainder of the {\it RXTE} mission.  \citet{Levine:2011a} failed
to reproduce it, but also did not use all the techniques that
\citet{Vanderlinde:2003a} used.  Associating it with an alias of the
true orbital period seems unlikely, as {\it RXTE}/ASM data on Sco~X-1
are rather well sampled through the year (just as ASAS data are).  

One possible explanation might be if the X-ray signal came at the beat
frequency between the orbital period and a superorbital period of
around a year.  Many X-ray binaries have indeed shown super-orbital
periods of tens to hundreds of days \citep[see e.g.][]{Charles:2008a},
although typically all are shorter than a year.  The only claim of
such a long period in Sco~X-1, came from early {\it RXTE}/ASM data,
from which \citet{Peele:1996a} suggested a 37\,day period.  This
detection has not been sustained in subsequent data, and no
super-orbital period was found by \citet{Farrell:2009a} in {\it
  Swift}/BAT data.  On longer timescales, \citet{Durant:2010a} and
\citet{Kotze:2010a} both independently suggested a $\sim9$\,year X-ray
modulation is present in {\it RXTE}/ASM data, although this is too
long to account for the \citet{Vanderlinde:2003a} period.  This
explanation therefore seems unlikely, and it remains to be seen if the
X-ray period can be reproduced from the full {\it RXTE} mission-long
dataset.

\section{Conclusions}

We have analyzed ASAS data of Sco~X-1 spanning nine years.  We can
confirm the period of \citet{Gottlieb:1975a}, while rejecting its
one-year aliases, and also the putative X-ray period of
\citet{Vanderlinde:2003a}.  Our updated ephemeris is $T_{\rm min}({\rm
  HJD}) = 2453510.329(17)+0.787313(1)E$.

\acknowledgments

This work was supported by the National Science Foundation under Grant
No. AST-0908789.  This research has made use of NASA's Astrophysics
Data System.

{\it Facilities:} \facility{ASAS}.


\begin{thebibliography}{}

\bibitem[Augusteijn et al.(1992)]{Augusteijn:1992a} Augusteijn, T.,
  Karatasos, K., Papadakis, M., et al.\ 1992, \aap, 265, 177

\bibitem[Bord et al.(1976)]{Bord:1976a} Bord, D.~J., Messina, 
R.~J., Mook, D.~E., \& Hiltner, W.~A.\ 1976, \apjl, 206, L49 


\bibitem[Charles et al.(2008)]{Charles:2008a} Charles, P., Clarkson, 
W., Cornelisse, R., \& Shih, C.\ 2008, New Astronomy Reviews, 51, 768 

\bibitem[Coe et al.(1980)]{Coe:1980a} Coe, M.~J., Dennis, B.~R., 
Dolan, J.~F., et al.\ 1980, \apj, 237, 148 


\bibitem[Cowley \& Crampton(1975)]{Cowley:1975a} Cowley, A.~P., \&
  Crampton, D.\ 1975, \apjl, 201, L65

\bibitem[Crampton et al.(1976)]{Crampton:1976a} Crampton, D., Cowley, 
A.~P., Hutchings, J.~B., \& Kaat, C.\ 1976, \apj, 207, 907 

\bibitem[Durant et al.(2010)]{Durant:2010a} Durant, M., Cornelisse, 
R., Remillard, R., \& Levine, A.\ 2010, \mnras, 401, 355 

\bibitem[Farrell et al.(2009)]{Farrell:2009a} Farrell, S.~A., Barret, 
D., \& Skinner, G.~K.\ 2009, \mnras, 393, 139 

\bibitem[Giacconi et al.(1962)]{Giacconi:1962a} Giacconi, R., Gursky, 
H., Paolini, F.~R., \& Rossi, B.~B.\ 1962, Physical Review Letters, 9, 439 

\bibitem[Gottlieb et al.(1975)]{Gottlieb:1975a} Gottlieb, E.~W., 
Wright, E.~L., \& Liller, W.\ 1975, \apjl, 195, L33 

\bibitem[Holt et al.(1976)]{Holt:1976a} Holt, S.~S., Boldt, E.~A., 
Serlemitsos, P.~J., \& Kaluzienski, L.~J.\ 1976, \apjl, 205, L27 

\bibitem[Kotze \& Charles(2010)]{Kotze:2010a} Kotze, M.~M., \&
  Charles, P.~A.\ 2010, \mnras, 402, L16

\bibitem[LaSala \& Thorstensen(1985)]{LaSala:1985a} LaSala, J., \&
  Thorstensen, J.~R.\ 1985, \aj, 90, 2077

\bibitem[Levine et al.(2011)]{Levine:2011a} Levine, A.~M., Bradt,
  H.~V., Chakrabarty, D., Corbet, R.~H.~D., \& Harris, R.~J.\ 2011,
  \apjs, 196, 6

\bibitem[Nelder \& Mead(1965)]{Nelder:1965a} Nelder, J.~A., Mead,
  R.\ 1965, The Computer Journal, 7, 308

\bibitem[Peele \& White(1996)]{Peele:1996a} Peele, A.~G., \& White,
  N.~E.\ 1996, \iaucirc, 6524, 2

\bibitem[Pojmanski(2002)]{Pojmanski:2002a} Pojmanski, G.\ 2002, Acta.\ Astron., 
52, 397 

\bibitem[Priedhorsky \& Holt(1987)]{Priedhorsky:1987a} Priedhorsky,
  W.~C., \& Holt, S.~S.\ 1987, \apj, 312, 743

\bibitem[Priedhorsky et al.(1995)]{Priedhorsky:1995a} Priedhorsky,
  W.~C., Brandt, S., \& Lund, N.\ 1995, \aap, 300, 415

\bibitem[Sandage et al.(1966)]{Sandage:1966a} Sandage, A., Osmer, P., 
Giacconi, R., et al.\ 1966, \apj, 146, 316 

\bibitem[Steeghs \& Casares(2002)]{Steeghs:2002a} Steeghs, D., \&
  Casares, J.\ 2002, \apj, 568, 273

\bibitem[Vanderlinde et al.(2003)]{Vanderlinde:2003a} Vanderlinde, K.~W., 
Levine, A.~M., \& Rappaport, S.~A.\ 2003, \pasp, 115, 739 

\bibitem[van 
Genderen(1977)]{vanGenderen:1977a} van Genderen, A.~M.\ 1977, \aaps, 28, 119 

\bibitem[Wright et al.(1975)]{Wright:1975a} Wright, E.~L., Gottlieb, 
E.~W., \& Liller, W.\ 1975, \apj, 200, 171 



\end{thebibliography}
\end{document}